\documentclass[12pt]{iopart}
\usepackage{graphicx}  
\begin{document}
\title{Structural, magnetic, and transport properties of Co$_2$FeSi Heusler films}

\author{H Schneider, Ch Herbort, G Jakob and H Adrian}
\address{Institut f\"ur Physik, Johannes Gutenberg-Universit\"at, 55099 Mainz, Germany}

\author{S Wurmehl and C Felser}
\address{Institut f\"ur Analytische Chemie und Anorganische Chemie, Johannes Gutenberg-Universit\"at, 55099 Mainz, Germany}

\ead{schneiho@uni-mainz.de}
\begin{abstract}
We report the deposition of thin $\rm Co_2FeSi$ films by RF magnetron sputtering. Epitaxial (100)-oriented and $\rm L2_1$ ordered growth is observed for films grown on MgO (100) substrates. (110)-oriented films on $\rm Al_2O_3\,(11\bar{2}0)$ show several epitaxial domains in the film plane. Investigation of the magnetic properties reveals a saturation magnetization of $\rm 5.0\,\mu_B/fu$ at low temperatures. The temperature dependence of the resistivity $\rho_{xx}(T)$ exhibits a crossover from a $T^{3.5}$ law at $T<50\,\mathrm{K}$ to a $T^{1.65}$ behaviour at elevated temperatures. $\rho_{xx}(H)$ shows a small anisotropic magnetoresistive effect. A weak dependence of the normal Hall effect on the external magnetic field indicates the compensation of electron and hole like contributions at the Fermi surface.
\end{abstract}

\maketitle

\section{Introduction}
Recently Co-based full-Heusler compounds have attracted a large research interest. Due to their high Curie temperatures and their predicted high spin polarisation they are prime candidates for the use in spinelectronic applications \cite{GAL02,FEC06}. In the fully ordered $\rm L2_1$ crystal structure these $\rm Co_2XY$ compounds consist of four interpenetrating fcc lattices where each sublattice is occupied by atoms of one element. Only for this structure full spin polarisation is predicted \cite{MIU06}. The growth of $\rm L2_1$ ordered Heusler films has been reported for example for $\rm Co_2MnGe$ \cite{AMB00} and $\rm Co_2Cr_xFe_{1-x}Ga$ \cite{UME05}, and magnetic tunneling junctions with epitaxial $\rm Co_2MnSi$ electrode show a high spin polarisation of 89\% at low temperatures \cite{OOG06}. Unfortunately this spin polarisation decreases rapidly with increasing sample temperature.

Another promising candidate for spintronic applications is $\rm Co_2FeSi$. Previous attempts in creating bulk samples of this Heusler alloy resulted in specimens with saturation magnetization values of $\rm 5.2\,\mu_B/fu$ \cite{BUS83} and $\rm 5.7\,\mu_B/fu$ \cite{NIC79}. From these values one would not expect $\rm Co_2FeSi$ to be a halfmetal, since these values are lower than expected from the Slater-Pauling rule for full Heusler alloys \cite{GAL02,WUR05}. However, recently fabricated bulk samples possess the expected value of $\rm 6\,\mu_B/fu$ for a halfmetallic system \cite{WUR05,WUR06}. Furthermore LDA+U calculations that reproduce this experimental fact yield full spin polarisation \cite{WUR05}. Additionally these bulk samples exhibit a Curie temperature of 1100~K. This is the highest value reported for Heusler alloys to date and might result in a smaller reduction of the spin polarisation at room temperatures. Epitaxial film growth and first tunneling junctions using A2-ordered electrodes have very recently been reported \cite{HAS05, INO06}. In this article we report the deposition and structural characterization of $\rm Co_2FeSi$ thin films on MgO (100) and $\rm Al_2O_3\,(11\bar{2}0)$ substrates. Furthermore we investigate the magnetic and transport properties of these films in order to gain insight in the electronic structure of these films.

\section{Film preparation and crystal structure}
\begin{figure}
\includegraphics[height=\columnwidth,angle=270]{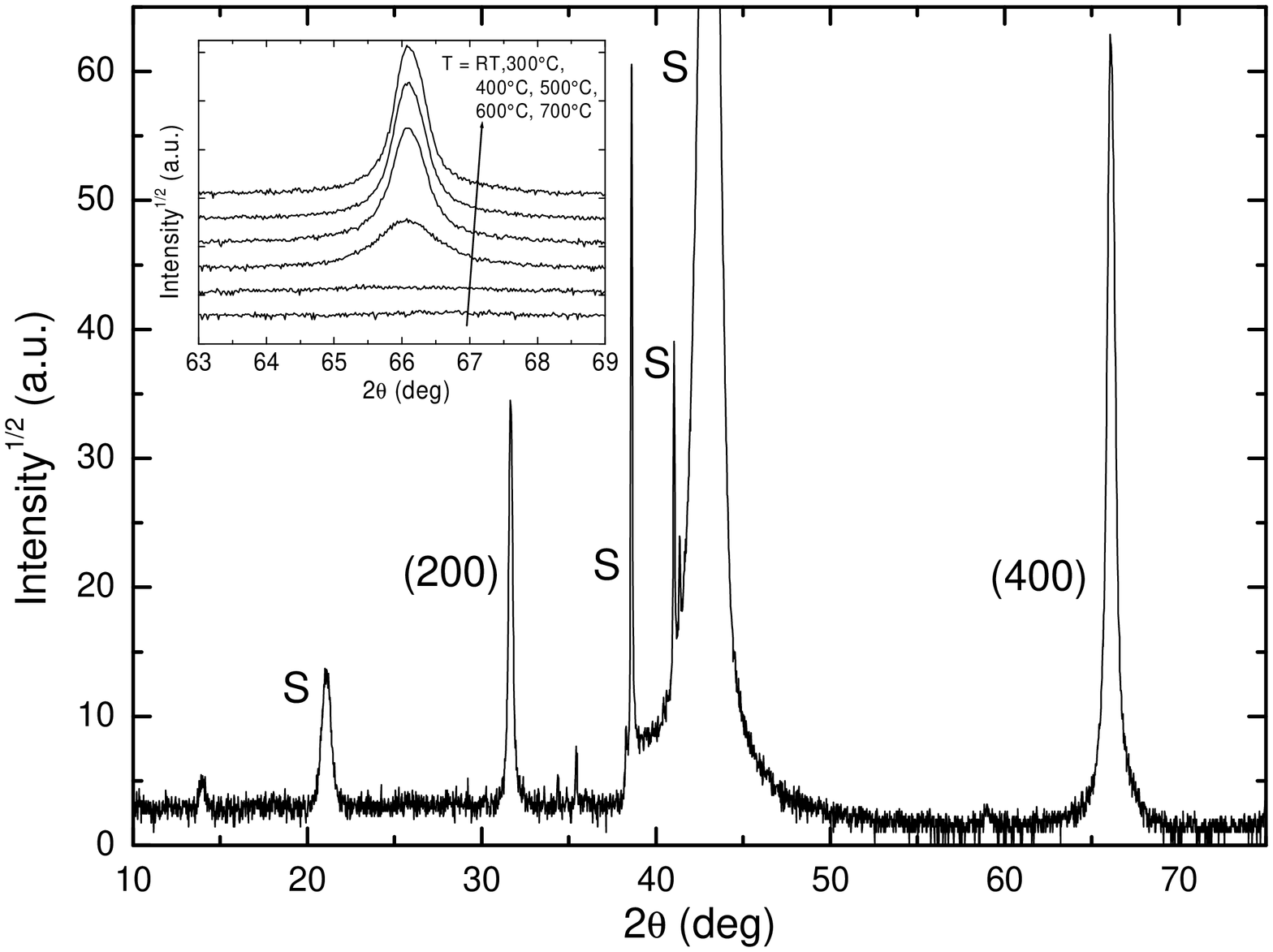}
\caption{Bragg scan of a $\rm Co_2FeSi/MgO$ film grown at a substrate temperature of 700$^\circ$C. The peaks marked with S are substrate reflections. The inset shows $\theta$-$2\theta$ scans of  the (400) reflection of films deposited at various substrate temperatures. The thickness of all films was 65~nm, for better visibility the scans are shifted along the ordinate.}\label{bragg}
\end{figure}

\begin{figure}
\includegraphics[width=\columnwidth]{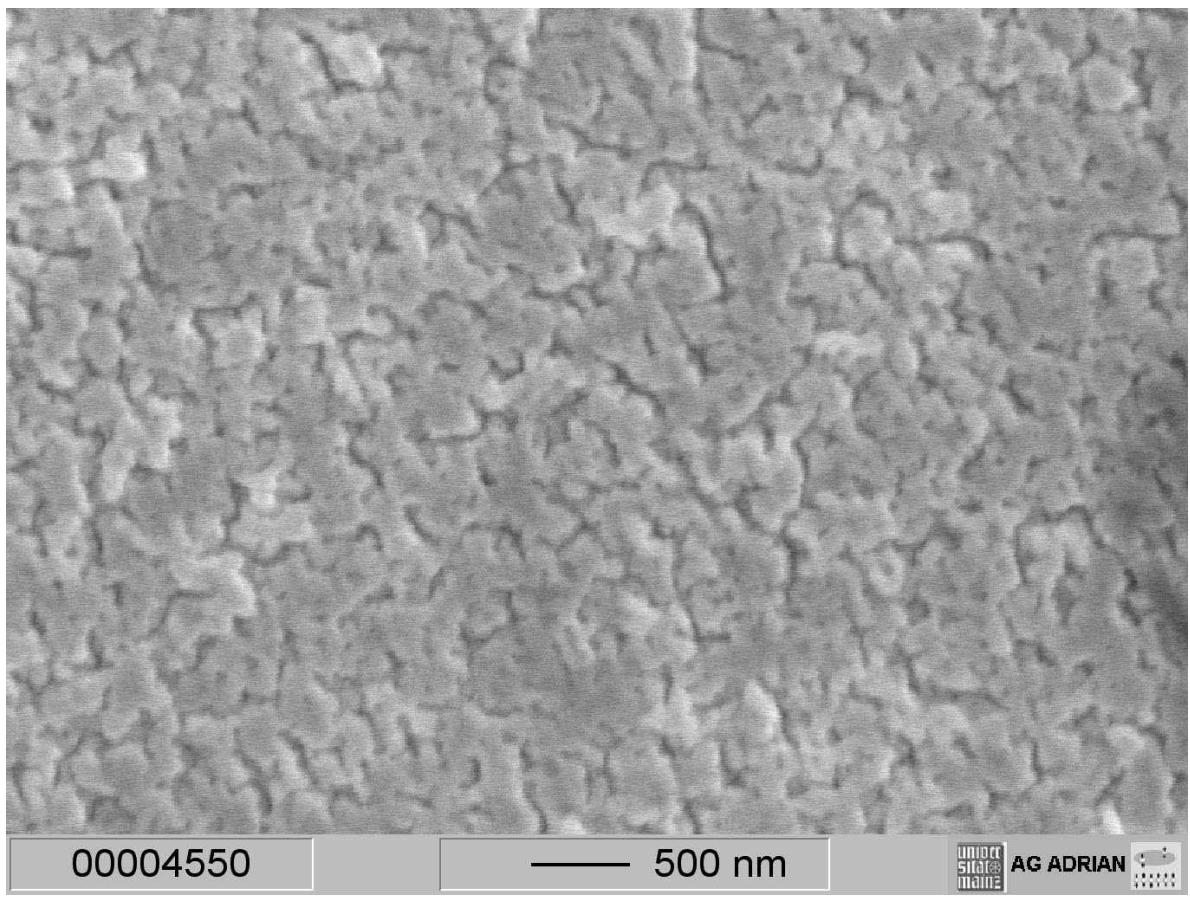}
\caption{SEM image of a $\rm Co_2FeSi/MgO$ film deposited at 700$^\circ$C (thickness = 80~nm).}\label{surface}
\end{figure}

\begin{figure}
\includegraphics[height=0.5\columnwidth,angle=270]{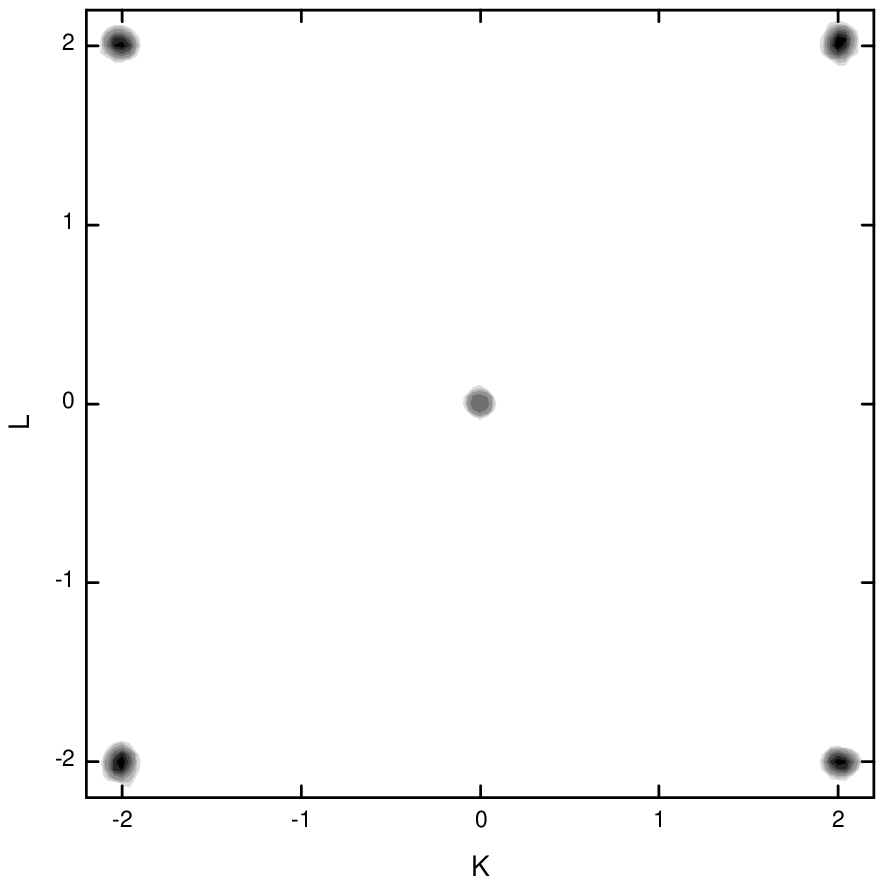}
\caption{Scan of the reciprocal (2KL) plane of $\rm Co_2FeSi$/MgO. The intensity is plotted with a logarithmic gray scale and a constant background is subtracted.}\label{fourcircle}
\end{figure}

\begin{figure}
\includegraphics[width=0.5\columnwidth]{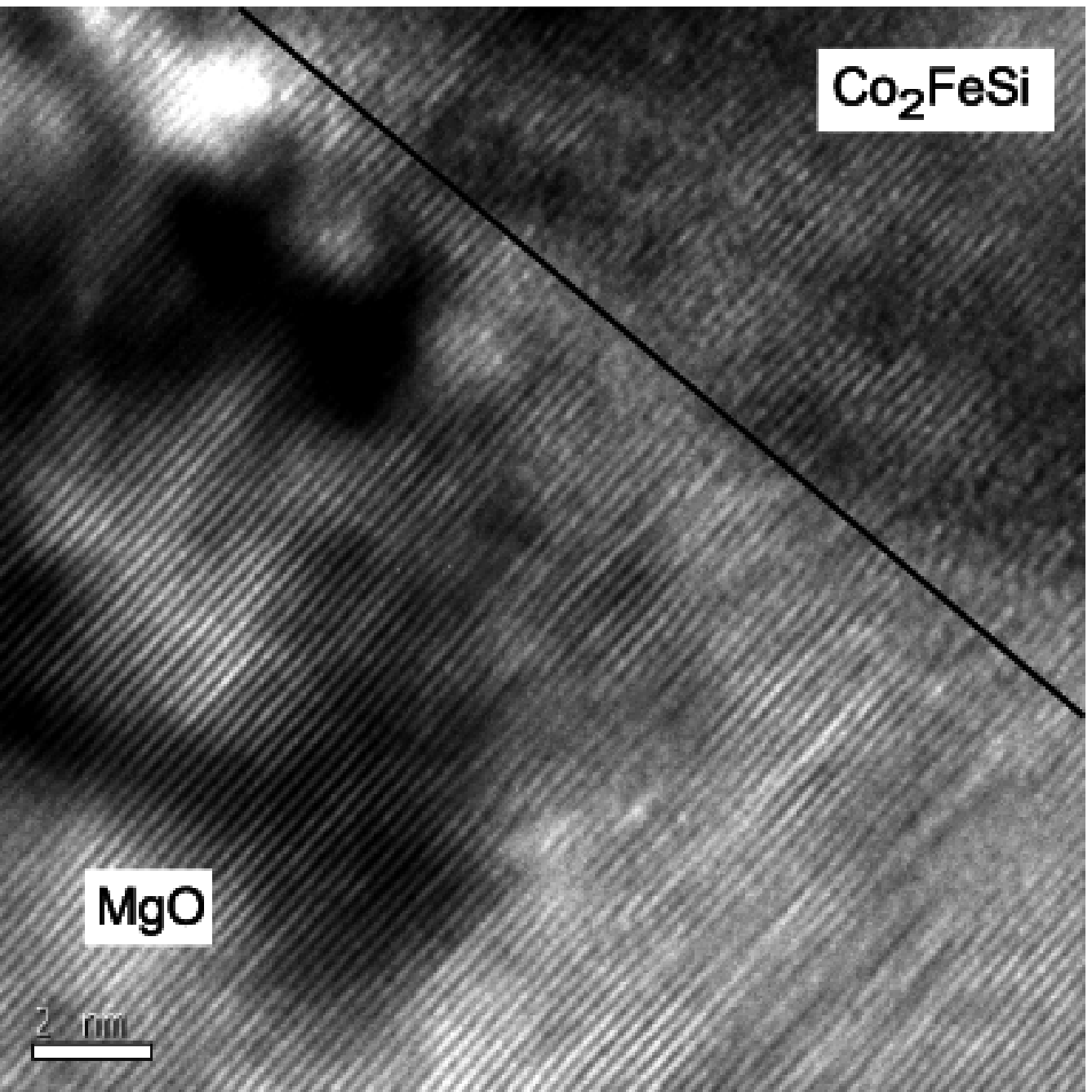}
\caption{HRTEM image of the interface between $\rm Co_2FeSi$ film and MgO substrate.}\label{tem}
\end{figure}

Thin films of $\rm Co_2FeSi$ were grown on $\rm Al_2O_3 (11\bar{2}0)$ and MgO (100) substrates by RF magnetron sputtering. The targets were cut from polycrystalline ingots which had been prepared by argon arc melting \cite{WUR05}. The base pressure was $\rm 5 \times 10^{-8}\, mbar$, the sputtering process was carried out in a flowing Ar atmosphere. Figure \ref{bragg} shows a Bragg scan of a 60~nm thick $\rm Co_2FeSi$ film grown on MgO (100) at a substrate temperature of 700$^\circ$C. The film grows (100)-oriented, the $\omega$-scan of the (400) reflection has a width of 0.3$^\circ$. The inset of Fig.~\ref{bragg} illustrates that for decreasing substrate temperatures the scattered intensities are reduced. This indicates that low deposition temperatures favour nanocrystalline rather than singlecrystalline growth. Subsequent annealing of these films did not improve the long-range crystal ordering of the samples. The differences between high temperature and low temperature growth are accompanied by a change of the surface morphology. Figure \ref{surface} shows an SEM image of a epitaxial film with a thickness of 80~nm. This film shows a flat surface pervaded with deep trenches. At lower depositions temperatures these trenches disapper.

Figure \ref{fourcircle} depicts a scan of the reciprocal (2KL) plane of a film deposited at 700$^\circ$C. Apart from the specular (200) reflection only the (220) and equivalent reflections are visible, which proves that these films grow fully epitaxial. The lattice constant has the bulk value of 5.64~\AA{}, the unit cell of the film is rotated by 45$^\circ$ with respect to the substrate. Furthermore the (111) and (311) reflections are present, which evidences $\rm L2_1$ ordered growth. By analyzing the relative intensities of these superlattice reflections we find that the $\rm L2_1$ ordering is not perfect, but a disorder between Si and Fe sites of 15 -- 20\% might be present \cite{SCH06}. 

Additional insight into the film growth process can be obtained from high-resolution transmission electron microscopy (HRTEM). The cross sections of our samples were prepared by mechanical thinning and subsequent Ar-ion polishing. Figure \ref{tem} shows an HRTEM image of a $\rm Co_2FeSi$ film deposited on MgO. Despite a lattice mismatch of 5.6\% between film and substrate only a slightly distorted growth directly at the interface is visible. After a few layers complete order is restored.

On $\rm Al_2O_3\,(11\bar{2}0)$ films grow (110)-oriented at substrate temperatures of 700$^\circ$C, the rocking curve of the (220) reflection has a width of 0.1$^\circ$. Again, the scattered intensity is reduced if the films are deposited on colder substrates. In the film plane, however, several epitaxial domains are observed. The preferred orientation is $\rm (1\bar{1}0)_{film}||(0001)_{sub}.$ The atomic site disorder is the same as in the films grown on MgO. The magnetic and transport properties presented in the following sections are identical for (100) and (110) oriented films unless stated otherwise.

\section{Magnetism}

\begin{figure}
\includegraphics[height=\columnwidth,angle=270]{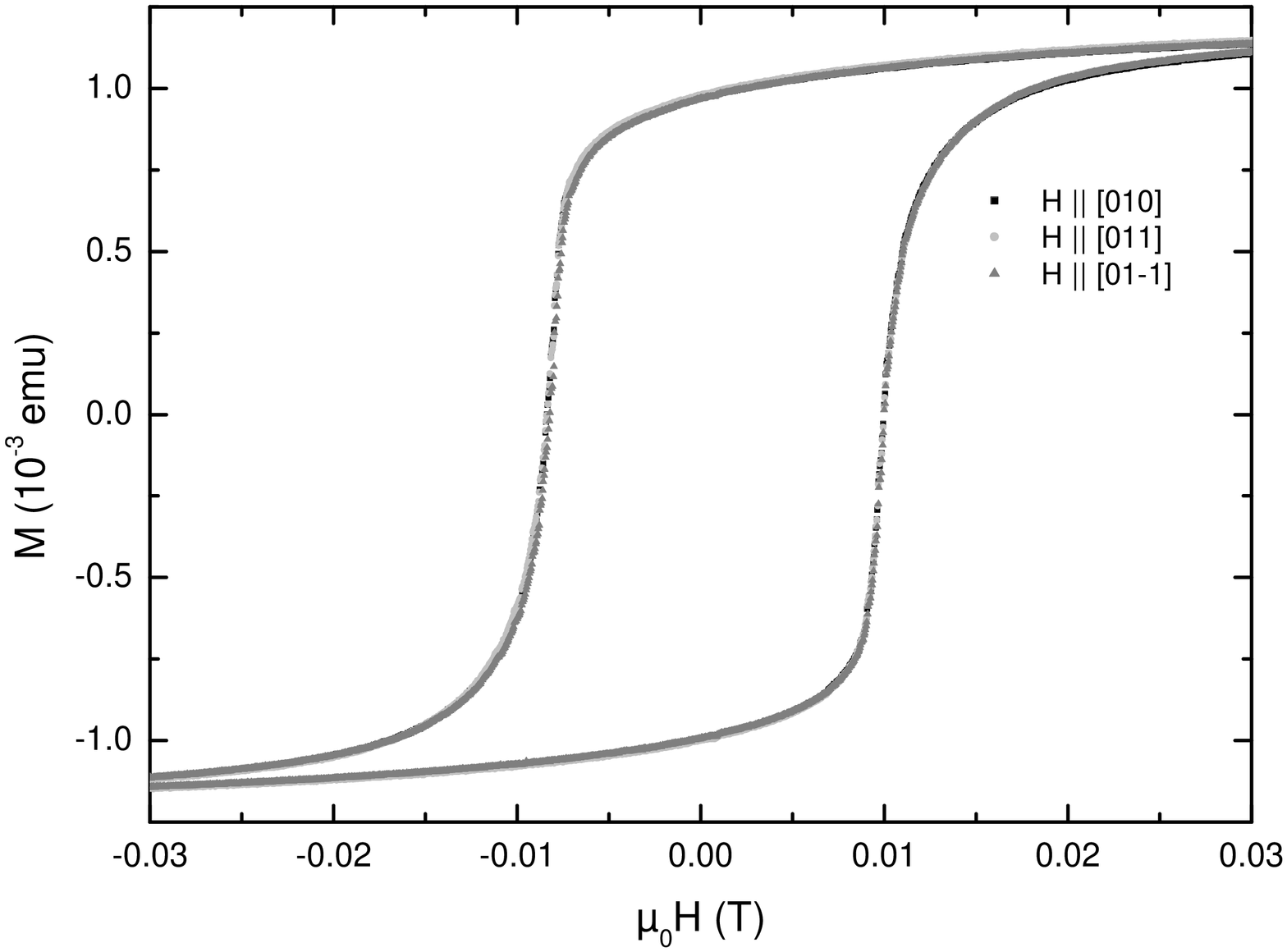}
\caption{Region of interest of hysteresis loops for different film orientations with respect to the external field. This $\rm Co_2FeSi/MgO$ film was deposited at 700~$^\circ$C, the measurements were performed at room temperature.}\label{vsm}
\end{figure}

\begin{figure}
\includegraphics[height=\columnwidth,angle=270]{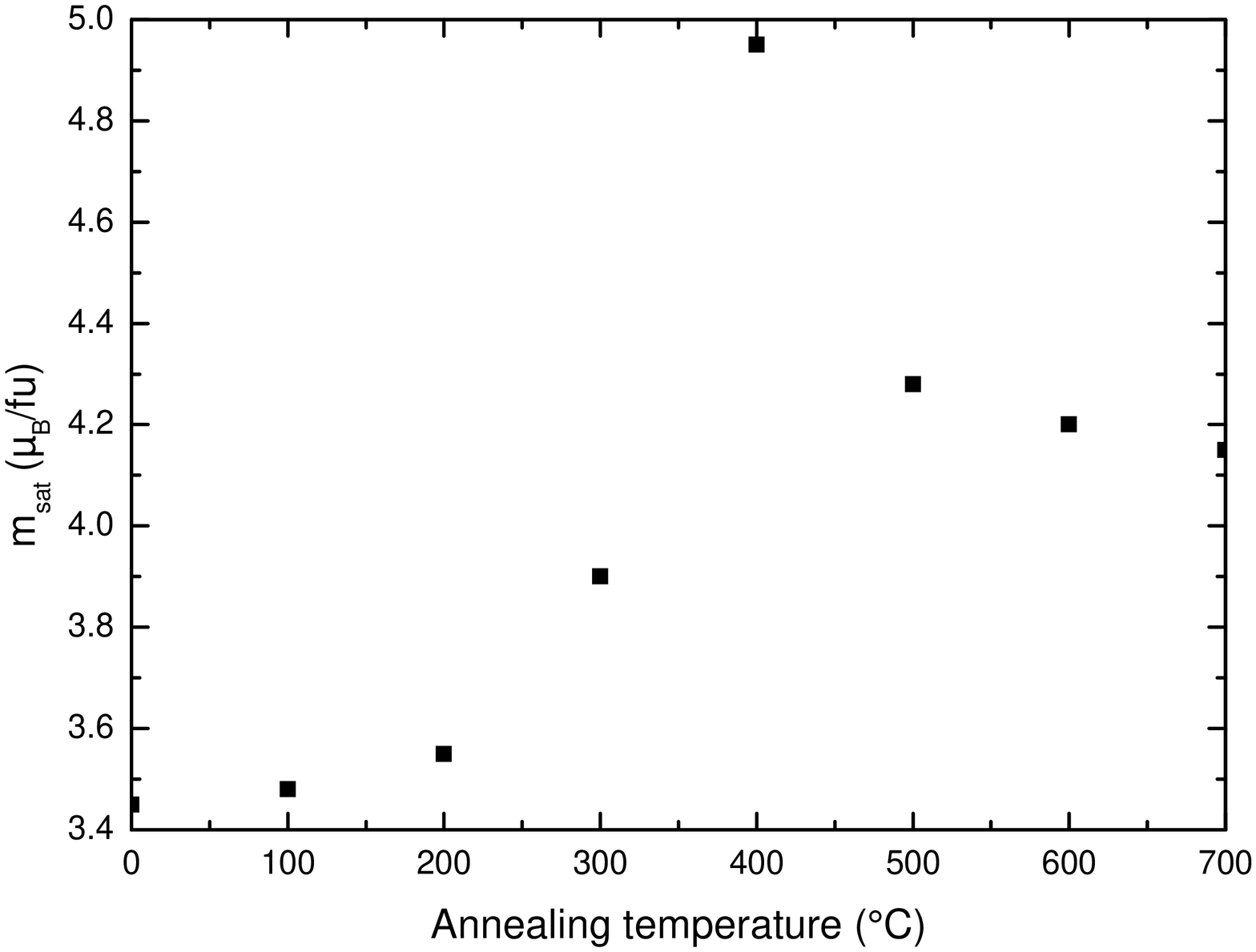}
\caption{Dependence of $m_{sat}(4\,\mathrm{K})$ of a film deposited at room temperature on annealing temperature.}\label{anneal}
\end{figure}

The magnetic properties of the films were investigated with a vibrating sample magnetometer. Room temperature magnetization loops of an epitaxial $\rm Co_2FeSi/MgO$ film along different directions are shown in Fig. \ref{vsm}. The measured volume magnetization corresponds to a saturation magnetization of $m_{sat}(300~\mathrm{K})=4.75\,\mu_B/\mathrm{fu}$. The temperature dependence of $m_{sat}$ follows a $T^{3/2}$ law upon cooling and can be extrapolated to $\rm 5.0\,\mu_B/fu$ at 0~K. This reduction of $m_{sat}$ compared to the bulk value is likely caused by the incomplete $\rm L2_1$ ordering of the crystal structure. The anisotropy of these films is extremely small, the difference in coercitivity is smaller than 1\%. This result was confirmed by measurements of the magneto-optical Kerr effect of the films \cite{HAM06}. In contrast, our films deposited on $\rm Al_2O_3$ as well as films deposited on GaAs (100) by molecular beam epitaxy \cite{HAS05} possess a uniaxial anisotropy with the easy axis along the $[1\bar{1}0]$ direction.
For films deposited at low temperatures we find a further reduction of $m_{sat}$. However, for these samples the value changes after annealing as illustrated in Fig. \ref{anneal}. Since no change in the scattered X-ray intensities is observed, we assume that although the films remain nanocrystalline, the annealing process alters $\rm L2_1$ ordering within the crystallites.

\section{Electronic transport}

\begin{figure}
\includegraphics[height=\columnwidth,angle=270]{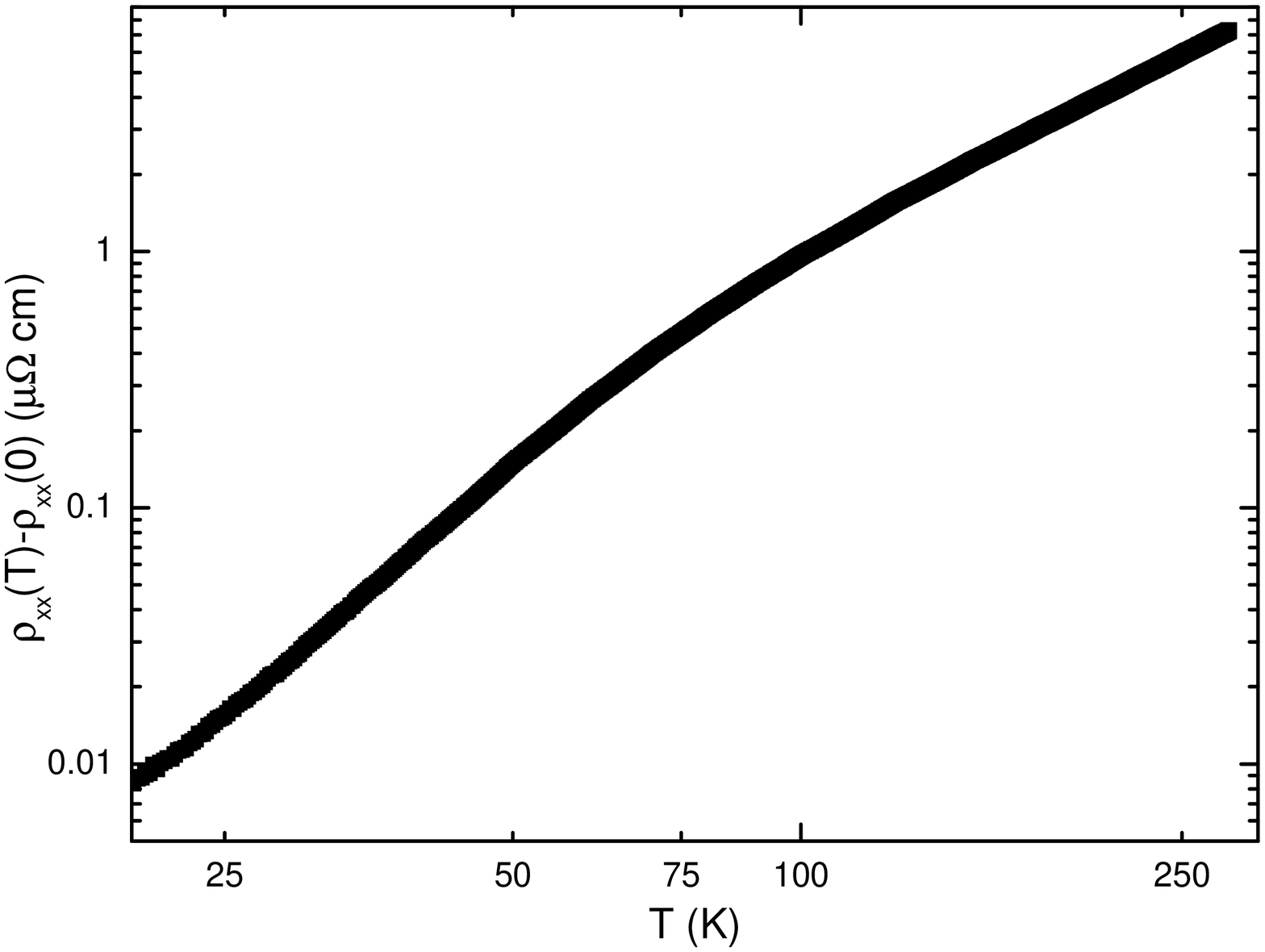}
\caption{Temperature dependence of the resistivity in zero field.}\label{resistivity}
\end{figure}

\begin{figure}
\includegraphics[height=\columnwidth, angle=270]{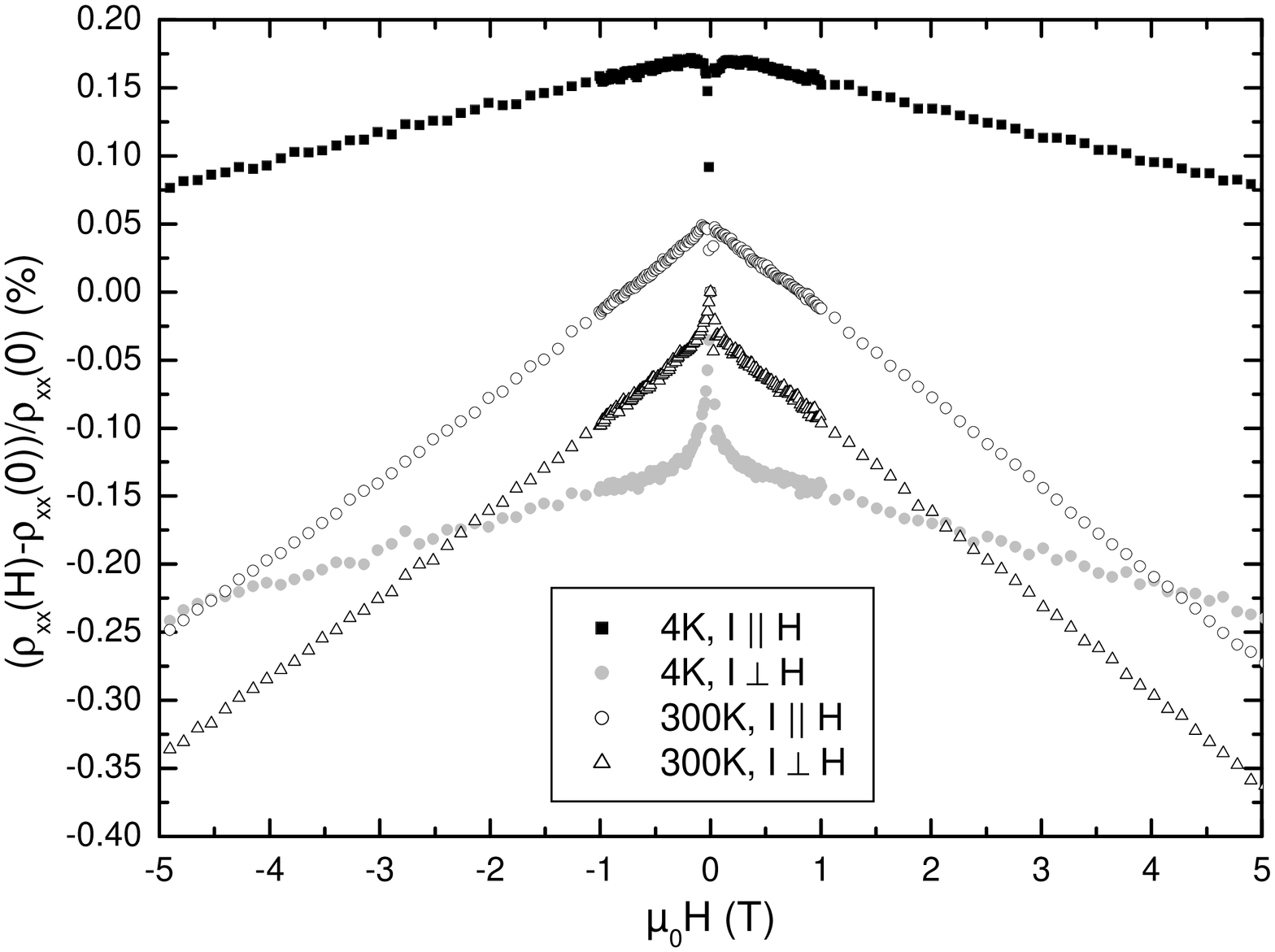}
\caption{Magnetoresistance of epitaxial $\rm Co_2FeSi$ thin film at 4~K and room temperature.}\label{mr}
\end{figure}

\begin{figure}
\includegraphics[height=\columnwidth,angle=270]{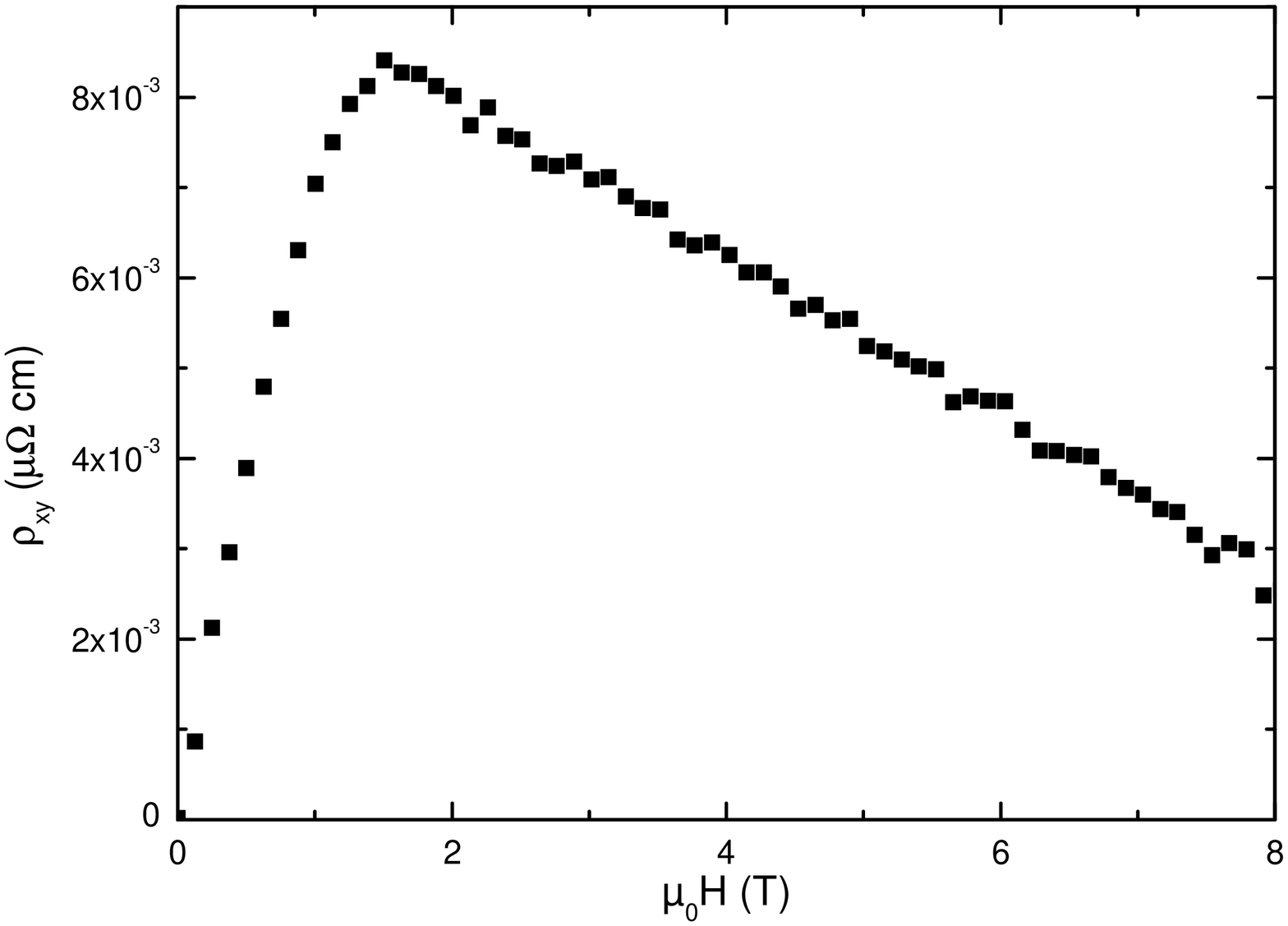}
\caption{Field dependence of the Hall resistivity at 4~K.}\label{hall}
\end{figure}

Despite the uneven surface of the epitaxial films they are electrically conducting down to a thickness of 5~nm. Patterning the samples by a standard photolitographic process with ion beam etching made it possible to obtain well defined geometries for the investigation of anisotropic transport properties. Resistivity and Hall effect were then measured using standard DC techniques. The temperature dependence of the film resistivity $\rho_{xx}(T)$ is shown in Fig. \ref{resistivity}. A residual resistance ratio of 1.5 and a residual resistivity of $\rm 30\,\mu\Omega cm$ was observed. This weak T-dependence is comparable to other thin films of Heusler alloys \cite{GEI02, SIN04} and is an indication for a strong contribution of impurity scattering to the resistivity. A $T^{3.5}$ dependence of $\rho_{xx}(T)$ at temperatures below 50~K is observed. As the sample temperature increases the exponent is reduced and reaches a value of 1.65 above 100~K. These exponents do not have a simple physical interpretation, since a number of effects can contribute to the temperature dependence of the resistivity of a ferromagnet, and the theoretical values for the exponents are usually obtained for simple systems only \cite{RIC79, GOO63, KUB72, FUR00}.

Figure \ref{mr} shows the magnetoresistive properties of our films. Below magnetic saturation we observe a small anisotropic magetoresistance effect. The spontaneous resistive anisotropy $(\rho_{||}-\rho_\perp)/(1/3\rho_{||}+2/3\rho_\perp)$ at 4~K has a value of 0.3\%. The effect decreases at higher temperatures, at room temperature a value of 0.08\% is observed. At higher fields we find a linear decrease of the resistivity. The slope increases from $\rm 6.5\,n\Omega cm/T$ at 4~K to $\rm 28\,n\Omega cm/T$ at room temperature. This behaviour has been observed in a number of ferromagnets and is generally attributed to a suppression of spin-flip scattering in high magnetic fields \cite{RAQ02}.

Hall effect data for $\rm Co_2FeSi$ at 4~K is presented in Fig. \ref{hall}. From the slope of the normal Hall effect we find a Hall constant of $R_H=8.9\times 10^{-12}\,\mathrm{m^3/As}$. This unusually small value corresponds to an  effective charge carrier density of 31 electrons/fu. It indicates the compensation of the Hall voltage by electron and hole like contributions from different parts of the Fermi surface.

\section{Conclusion}
We have successfully sputtered thin epitaxial films of the full Heusler compound $\rm Co_2FeSi$. They grow in the ordered $\rm L2_1$ structure. However, some disorder was found in these films. This disorder causes a reduction of the saturation magnetization compared to the values found in bulk samples and expected from the Slater-Pauling rule. It is also perceivable in a weak temperature dependence of the electrical resistivity. The analysis of the magnetoresistive data reveals the presence of an anisotropic magnetoresistive effect as well as the presence of spin-flip scattering in zero field even at low temperatures. From these properties we have to conclude that the very good, but not perfect, crystal structure of these Heusler films leads to a loss of the expected halfmetallicity and an improved deposition process, for example the use of a buffer layer, is required.

Financial support by the DFG (Forschergruppe 559) is acknowledged.

\end{document}